%% file: io_control_dataset.tex
\def\year{2022}\relax
\documentclass[letterpaper]{article} 
\usepackage{aaai22}  
\usepackage{times}  
\usepackage{helvet}  
\usepackage{courier}  
\usepackage[hyphens]{url}  
\usepackage{graphicx} 
\usepackage{natbib}  
\usepackage{caption} 
\DeclareCaptionStyle{ruled}{labelfont=normalfont,labelsep=colon,strut=off} 
\usepackage{algorithm}
\usepackage{algorithmic}
\usepackage{multirow}
\usepackage{booktabs}
\usepackage{tabularx}

\usepackage{xcolor}
\newcommand{\answerYes}[1]{\textcolor{blue}{#1}} 
\newcommand{\answerNo}[1]{\textcolor{teal}{#1}}

\usepackage{newfloat}
\usepackage{listings}
\floatstyle{ruled}
\newfloat{listing}{tb}{lst}{}
\floatname{listing}{Listing}
\title{Labeled Datasets for Research on Information Operations}

\author{
    Ozgur Can Seckin\thanks{Corresponding author. Email: \texttt{oseckin@iu.edu}.\\$^\dag$Equal contributions.}$^\dag$\textsuperscript{\rm 1},
    Manita Pote$^\dag$\textsuperscript{\rm 1},
    Alexander C. Nwala$^\dag$\textsuperscript{\rm 2},
    Lake Yin\textsuperscript{\rm 1},
    Luca Luceri\textsuperscript{\rm 3},
    \\Alessandro Flammini\textsuperscript{\rm 1},
    Filippo Menczer\textsuperscript{\rm 1}
}

 \affiliations{
     \textsuperscript{\rm 1}Observatory on Social Media, Indiana University, Bloomington\\
     \textsuperscript{\rm 2}William \& Mary, Williamsburg, Virginia\\
     \textsuperscript{\rm 3}University of Southern California, Information Sciences Institute, Los Angeles\\
 }

\begin{document}

\maketitle

\begin{abstract}
Social media platforms have become a hub for political activities and discussions, democratizing participation in these endeavors. 
However, they have also become an incubator for manipulation campaigns, like information operations (IOs).
Some social media platforms have released datasets related to such IOs originating from different countries. However, we lack comprehensive control data that can enable the development of IO detection methods.
To bridge this gap, we present new labeled datasets about 26 campaigns, which contain both IO posts verified by a social media platform and over 13M posts by 303k accounts that discussed similar topics in the same time frames (control data).
The datasets will facilitate the study of narratives, network interactions, and engagement strategies employed by coordinated accounts across various campaigns and countries. By comparing these coordinated accounts against organic ones, researchers can develop and benchmark IO detection algorithms.
\end{abstract}

\section{Introduction}
\label{sec:introduction}

Originally perceived as tools to democratize information, social media platforms have also evolved into channels for the dissemination of conspiracy theories and questionable information \cite{Lazer-fake-news-2018, vosoughi2018spread}. 
The proliferation of inauthentic accounts \cite{Shao18hoaxybots,yang2023anatomy}, political sock puppets \cite{woolley2018computational}, and state-sponsored operators \cite{badawy2019characterizing} exacerbates the vulnerability of social media to misleading narratives and propaganda. 
Deceptive, orchestrated campaigns, known as \textit{information operations} (IOs), have been defined as coordinated efforts to manipulate or corrupt public debate within a target audience for a strategic goal \cite{facebook2021threatreport}.

Accounts participating in IOs employ various strategies, ranging from artificial amplification for promoting content to targeted attacks on specific account communities. IOs can be complex, sophisticated efforts characterized by several elements, including:
\begin{enumerate}
  \item \textbf{Domains}: The primary focus can vary widely across operations, but in most cases it is political.
  \item \textbf{Goals}: IOs can encompass a wide range of objectives, such as advancing narratives about politics, amplifying pro- or anti-government content, and spreading propaganda and/or disinformation. 
  \item \textbf{Targets}: The audiences subject to manipulation. These can vary in scale, ranging from small groups to entire countries and geopolitical regions.
  \item \textbf{Tactics}: The methods employed to achieve an IO's strategic goals. These tactics can range from simple actions (e.g., spamming) to complex ones involving coordination, obfuscation, or impersonation of political figures.
  \item \textbf{Platforms}: The activities associated with  IOs can extend across multiple online platforms.
  \item \textbf{Users}: The accounts involved in IOs may range from small to large sets of accounts, human-operated or automated, and may involve fake or compromised profiles.
\end{enumerate}

One of the first documented IOs was uncovered in a South Korean social media platform in 2012 \citep{keller2020political}. 
Since then, IOs have been reported globally and have emerged as a global threat \cite{bradshaw2017troops, woolley2019conclusion, sio}.
A well-known case is the interference in the 2016 U.S. Presidential Election by the Russian Internet Research Agency (IRA) \citep{senate_intelligence_committee_2019}. 
As a result of the potential adverse effects of IOs, major social media platforms like Twitter (now X), Facebook, and Reddit started releasing reports and data on IOs identified on their platforms. 

Research on IOs has mainly utilized Twitter data to characterize IO accounts, uncover their tactics, and propose methods for their detection \citet{nwala2023language,luceri2024unmasking,cima2024coordinated,saeed2024unraveling}. 
This research requires control data that include the activity of a baseline or negative class, i.e., legitimate accounts engaged in conversations similar to those of IO campaigns. 
The availability of control datasets for multiple IO campaigns is critical for at least two reasons. First, existing datasets are obsolete, private, or tied to a specific campaign, hindering the development of detection models that can generalize across IOs with varying origins, contexts, and levels of sophistication \cite{badawy2019characterizing, cima2024coordinated}. 
Second, following the shutdown of the Twitter API \cite{murtfeldt2024rip}, it has become prohibitively expensive for researchers to collect new data related to IOs. 
Nevertheless, control datasets for multiple IO campaigns remain underdeveloped. 
To bridge this gap, here we introduce new datasets that include both IO data and related control data covering legitimate accounts involved in online discussions across 26 distinct, verified IO campaigns from different countries. 
Our datasets are anonymized to preserve user privacy. 
Such comprehensive, accessible, and contextually enriched datasets provide an invaluable resource for researchers to analyze and characterize IOs in various contexts, as well as develop new methodologies to detect and counter IOs.

The datasets are available 
at https://doi.org/10.5281/zenodo.14141549.

\section{Related Work}

\subsection{Characterizing IOs and Their Tactics}

Previous research has explored the activity of IO accounts, such as state-sponsored trolls targeting the \#BlackLivesMatter movement \cite{stewart2018examining} and the 2016 U.S. Election \cite{badawy2019characterizing}, and the differences in their activities across campaigns  \cite{zannettou2019let}. 
Some studies have shown how IO accounts leverage inauthentic or automated accounts to increase their prominence and artificially amplify messages \cite{linvill_warren_2020, elmas2023analyzing}, while being resilient to large-scale shutdown \cite{Merhi2021}. 
Researchers have reported on different tactics used by IO accounts, such as trolling \cite{zannettou2019disinformation}, flooding through political cartoons \cite{fecher2022oh}, hashtag hijacking  \cite{ong2018architects}, deletion of content to avoid detection \cite{Torres2022deletions}, disinformation and propaganda \cite{woolley2019conclusion}, political memes \cite{rowett2018strategic, zannettou2020characterizing, ng2022coordinated}, and advertising or paid digital influencers \cite{ong2018architects}. 

\subsection{Detecting Inauthentic Coordinated Behaviors}

A variety of unsupervised and supervised machine-learning models have been developed ntify messages and coordinated accounts linked to IOs.
Unsupervised methods to detect coordination include multi-view modularity clustering \citet{uyheng2022mapping}, a Bayesian approach based on narrative and account characteristics \cite{smith2024unsupervised}, and network-based models analyzing similarities in sharing activities \cite{pacheco2021uncovering,luceri2024unmasking,cima2024coordinated}.
Some supervised learning approaches classify posts to determine if a message is part of an IO. These methods employ linguistic features to train off-the-shelf machine learning algorithms \cite{addawood2019linguistic, im2020still} or large language models \cite{luceri2023leveraging}. 
Other classifiers attempt to differentiate between IO accounts and organic accounts based on their behaviors. 
Methods for this task have employed inverse reinforcement learning \cite{luceri2020detecting}, sequences of account actions \cite{nwala2023language,ezzeddine2023exposing}, and Hawkes modeling~\cite{kong23Interval-censored}.  
Coordinated account detection has also exploited content \cite{alizadeh2020content-based} and generative models \cite{sharm2021hidden}. 
Recent machine-learning models have leveraged features extracted from cross-campaign data, such as third-party applications and reposting patterns, to detect accounts from previously unseen campaigns \cite{saeed2024unraveling}. 

\subsection{Collecting IO Control Datasets}

Combating information operations requires accessible data about inauthentic coordinated activity. Early examples of such data appeared after 2016, when Congress investigated Russian interference in the U.S. Election. The investigation uncovered malicious trolls and bots linked to the IRA that spread biased information, prompting social media platforms to address coordinated and fake activities. 
The House Intelligence Committee released redacted PDF files of 3,517 Facebook ads bought by the IRA.\footnote{\url{github.com/simonw/russian-ira-facebook-ads-datasette}} 
Meta also shared blog posts and threat reports to inform users about its efforts against misinformation.\footnote{\url{about.fb.com/news/tag/coordinated-inauthentic-behavior/}} 
Reddit identified 944 suspicious accounts possibly linked to the IRA\footnote{\url{www.reddit.com/wiki/suspiciousaccounts/}} along with a transparency report.\footnote{\url{www.reddit.com/r/announcements/comments/8bb85p/reddits_2017_transparency_report_and_suspect/}}

Although these datasets offer a good descriptive understanding of coordinated behavior, they fall short of helping differentiate between coordinated and regular accounts. Researchers have collected control datasets for specific campaigns released by Twitter, allowing for a more nuanced comparison between these groups. 
For example, \citet{badawy2019characterizing} collected tweets based on a list of hashtags and keywords related to the 2016 U.S. Presidential Election. 
\citet{alizadeh2020content-based} curated control data for multiple IO campaigns by combining random account IDs and accounts who followed at least five American politicians. 
\citet{vargas2020detection} collected four control groups for IO campaigns in 2018--2019: a political community (U.S. Congress and UK Parliament members), a non-political community (academic and security researchers), a group based on a trending hashtag, and popular accounts selected by a random walk through the follower network.
\citet{smith2024unsupervised} curated control datasets for four campaigns by querying specific keywords associated with each campaign within the same period. 
\citet{cima2024coordinated} compiled a control dataset for two IO campaigns. They selected the top hashtags used by the coordinated accounts in each campaign and collected all tweets that included at least one of these hashtags during the last 4 months of the campaigns. 
Similarly, \citet{guo2022large} introduced control datasets for 28 IO campaigns from 14 countries and spanning 2015--2018. The authors included the tweets containing the top hashtags used by the coordinated accounts in each month and gathered the tweets from a 1\% sample of real-time tweets provided by the Internet Archive.\footnote{\url{archive.org/details/twitterstream}}  

While existing control datasets are valuable, they have significant limitations we aimed to address when curating the control datasets presented in this paper.
First, they typically include control accounts who have posted on similar topics to those of the IO accounts, but they fail to include posts from these control accounts on unrelated topics. 
In contrast, our datasets include posts from control accounts discussing similar topics as well as other posts from their timelines, providing a more comparable set of posts to those of IO accounts.
Second, while most previous work provides control datasets for only a small number of IO campaigns and countries, our datasets include data for 26 campaigns from multiple countries, involving operations backed by several states. 
Third, some existing datasets were assembled using a small sample of public posts. Our datasets have 100\% coverage of control data. 
Finally, most previous datasets provide only IDs and need to be ``re-hydrated'' using platform APIs that are currently inaccessible to researchers. We provide anonymized data that complies with platform privacy policies and does not need require re-hydration. 

\section{IO Datasets}
\label{sec:twitter-io-dataset}

We curated control datasets for a number of IO datasets. 
We started from a public archive containing data on state-sponsored IOs, made available by a social media platform on their transparency website. 
The campaigns spanned several years and countries. 

After taking down the IO campaigns, the platform released corresponding datasets. We focus on 26 of the campaigns, attributed by the platform to 16 state actors. Each campaign is identified by a state actor (e.g., Russia or Catalonia) and a number to distinguish campaigns from the same state actor. 
A campaign contains records about the entire timelines of the IO accounts associated with the campaign --- whether or not each post in these timelines is part of the campaign. 
For example, a hijacked account might have been repurposed to become part of a campaign. 
For each campaign, we gathered these records from multiple files after consulting documentation by the platform, in the form of a README file and blog post, and cross-referencing the numbers of IO accounts.

We found four groups of campaigns that could be merged according to platform documentation:
\begin{enumerate}
    \item \textbf{Venezuela\_1}
    and \textbf{Venezuela\_2}
    
    \item \textbf{China\_1} and \textbf{China\_2}

    \item \textbf{Russia\_1}
    and \textbf{Russia\_4}
    
    \item \textbf{Iran\_2}, \textbf{Iran\_3}, and \textbf{Iran\_4}
\end{enumerate}
However, we deliberately refrained from merging these datasets, leaving it for researchers to do so if necessary or beneficial for their investigations.

\section{Control Datasets}
\label{sec:control-data-collection}

What constitutes a good control dataset for a given IO campaign is debatable. 
One of the main tactics of IO actors is to coordinate efforts to get engagement, interaction, and trust from organic accounts, who can be both targets and unwitting collaborators in pushing their agenda \cite{starbird2019disinformation}.
However, not every organic account is susceptible to IO influence attempts. Some can discuss similar topics without endorsing the IO messages. 
Therefore, a suitable control would consist of accounts who engage in the same topics at the same time, but who are not part of the IO.
One method to capture such topics is by extracting hashtags used in the IOs.

\subsection{Control Data Collection}

\begin{figure*}[t]
    \centering
    \includegraphics[width=\textwidth]{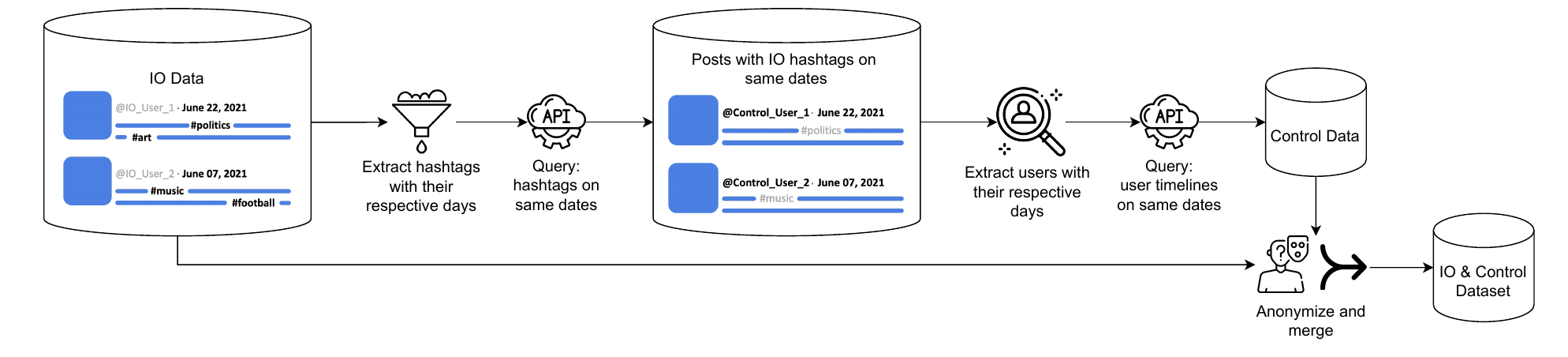}
    \caption{Data Collection and Curation Pipeline}
    \label{fig:control_data_curation}
\end{figure*}

Fig.~\ref{fig:control_data_curation} illustrates our data collection and curation pipeline.  
We first gathered all the hashtags used by IO accounts in a given IO campaign to identify control accounts discussing similar topics. 
Note that this approach is followed separately for each of the campaigns, e.g., the control datasets for \textbf{China\_1} and \textbf{China\_2} were collected independently.
Subsequently, we utilized these hashtags as queries to identify accounts that had posted on the same dates and used the same hashtags as the IO accounts, using the platform's Application Program Interface (API).
In the final step, we reconstructed the daily timelines of the control accounts by extracting up to 100 messages posted on the same dates as the IO accounts. For instance, if an IO account posted with the hashtag \texttt{\#election} on September 10, 2019, we created the corresponding control account list by identifying accounts who also used that hashtag on the same date. We then pulled their timelines from the API for that specific date.
We collected control data for each of the 26 IO campaigns. Our selection of these campaigns was in part dictated by API and computational limitations. 

\subsection{Data Curation}
\label{data-preprocessing}

We removed from the IO data the columns without corresponding fields in the control data from the API. Similarly, we removed from the control data any fields without corresponding columns in the IO data. 
Finally, we aligned the field names to ensure consistency between IO and control data. 

The merged datasets have the following fields:
\textit{postid, post\_text, application\_name, post\_language, in\_reply\_to\_postid, in\_reply\_to\_accountid, post\_time, accountid, account\_profile\_description, follower\_count, following\_count, account\_creation\_date, is\_repost, reposted\_accountid, reposted\_postid, hashtags, urls, account\_mentions, is\_control}. 
Records from IO campaigns are marked as \texttt{False} in the \textit{is\_control} column, while control records are marked as \texttt{True}. 

Before making the dataset publicly accessible, it is imperative to honor the privacy of users. Therefore, we anonymized all personally identifiable information (PII) through a one-way hashing algorithm. This includes account IDs, post IDs, URLs and usernames. The last two items are hashed, even when occurring within post text and profile description. Location information is removed.
We ensured consistent anonymization so that accounts mentioned in IO and control posts can be linked through their hashed IDs while still safeguarding privacy. 

Each dataset was segmented into files of 50,000 posts each. This resulted in a total of 703 files across all campaigns, with the largest campaign (\textbf{Cuba}) consisting of 124 files. 

\subsection{Descriptive and Coverage  Statistics}
\label{sec:data-coverage}

Table~\ref{tab:descriptive-stats} provides a quantitative overview of the datasets. The dataset with the longest duration spans 12 years (\textbf{China\_2}), while the shortest lasts less than a year (\textbf{Spain} and \textbf{Iran\_5}). Note that the first post in a dataset does not necessarily represent the beginning of a campaign, since IO accounts could exist and post before they were involved in a campaign.

Account and post counts vary significantly from one campaign to another, as do the numbers of reposts and replies.  This indicates that the levels of engagement differ across campaigns. 
We also observe that the number of control accounts always exceeds that of IO accounts, while the activity per IO accounts tends to be higher. 
Note, however, that this is likely a consequence of the different way in which the control data was collected --- individual days rather than full timelines. 

\input{tables/descriptive_statistics}

\begin{figure*}[t]
    \centering
    \includegraphics[width=\textwidth]{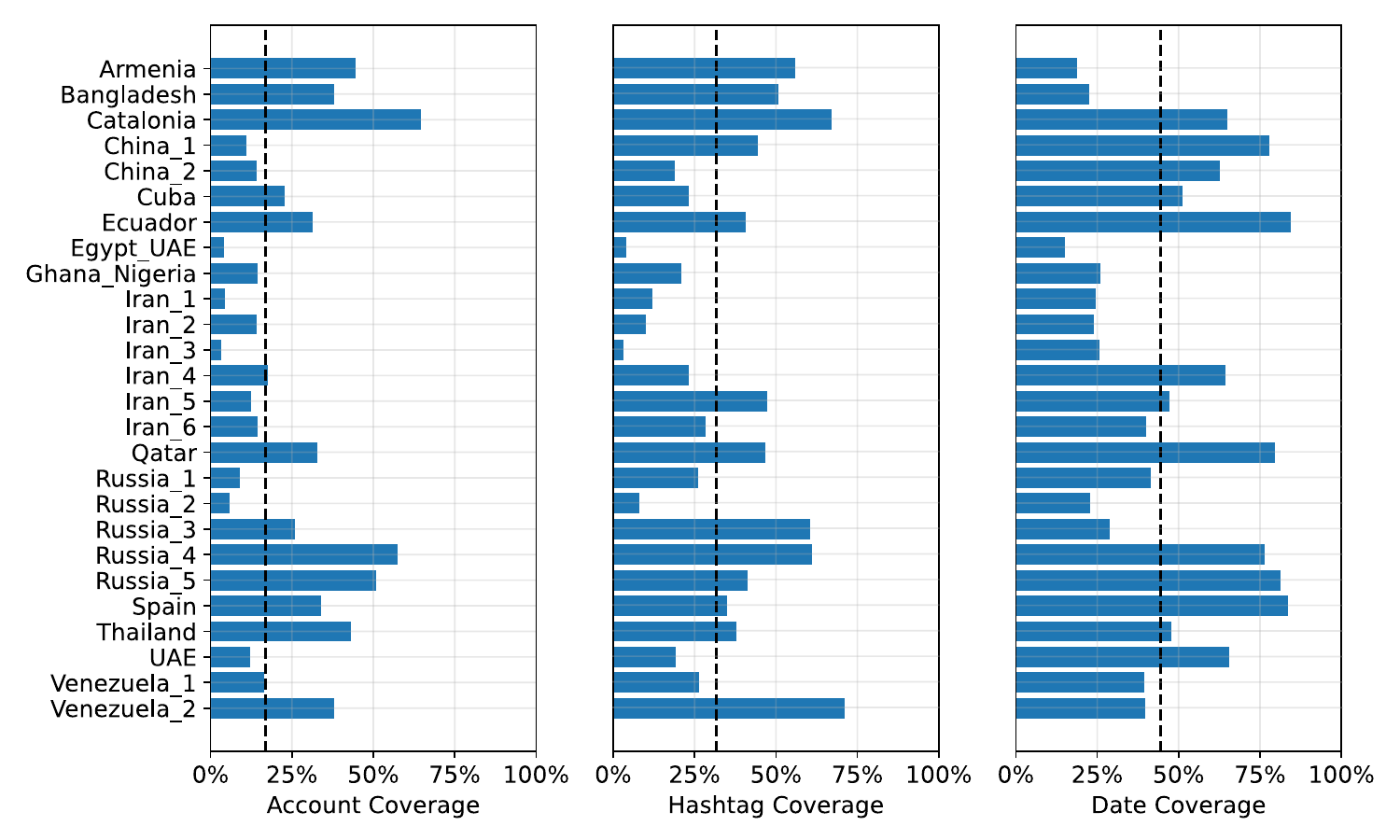}
    \caption{Data Coverage. Left: Percentage of accounts mentioned/replied to/reposted by IO accounts that can be found in the control dataset. Middle: Percentage of IO hashtags that were also used by control accounts. Right: Percentage of days that IO accounts posted and that were covered by control accounts. Dashed lines indicate median values.}
    \label{fig:data_coverage}
\end{figure*}

Fig.~\ref{fig:data_coverage} reports on the coverage of IO campaigns by control data in terms of accounts, hashtags, and time.
Account coverage refers to the portion of accounts mentioned, replied to, or reposted by IO accounts who were also mentioned, replied to, or reposted by control accounts. 
Account coverage varied between 3\% (\textbf{Iran\_3}) and 58\% (\textbf{Catalonia}), with a median of 17\%. 
The median coverage of IO hashtags was 31\%, with a range from 3\% (\textbf{Iran\_3}) to 73\% (\textbf{Venezuela\_2}).
Finally, the median date coverage of IO data was 44\%, ranging from 15\% (\textbf{Egypt\_UAE}) to 84\% (\textbf{Ecuador}).

\section{Discussion}
\label{sec:discussion}

Researchers can leverage the datasets introduced in this paper to characterize different dimensions of inauthentic coordinated activity. 
This includes analyzing differences between the narratives, behavioral patterns, network structures, and temporal activities of IO accounts, across campaigns and countries. 
Researchers can also investigate the tactics employed by different campaigns, for example how IO accounts engage with their targets. 
In addition, the datasets contribute systematic control data for IO campaigns, empowering researchers to design and evaluate methods for their detection. 

There are some limitations in the way the control data was collected. 
The choice to condition the collection of control accounts on the co-sharing of IO hashtags means that the quality of the control sample depends on IO hashtags being a quality proxy for IO content, which is not always the case. 
For example, some IO accounts may have employed tactics that did not involve the use of any hashtags. In such cases, our datasets may lack accurate control counterparts for those accounts. 

Even when a campaign does actively use hashtags, some of the hashtags may be very popular and generic. 
Using such hashtags to select control accounts can introduce a sampling bias towards more active accounts. 
Because of the inclusion of popular hashtags, control data may have non-IO coordinated activity, which would be inaccurately labeled in the dataset. 
As an example, suppose an IO account used the hashtag \texttt{\#crypto}. The control data may include accounts managed by a spammer to push cryptocurrency manipulation. While these accounts would be engaged in coordinated inauthentic activity, they would not be labeled as IO in the dataset, therefore an evaluation would wrongly mark them as false positive errors if they were correctly identified by a detection algorithm.

While the entire user timeline is present for IO accounts, control account timelines are cropped at 100 posts and exclude posts that occurred after the date on which they met the inclusion criteria. This results in a significant misalignment in the active days for individual control accounts compared to individual IO accounts. Although the posts by control accounts have reasonable \textit{collective} coverage of the IO activity dates (Fig.~\ref{fig:data_coverage}), a \textit{single} control account timeline tends to cover relatively short periods while the IO account timelines can span years. 
Research who wish to mitigate the effect of this bias might selectively choose IO and control accounts with matching temporal activity.

In addition to temporal bias, our coverage of IO and control hashtag use is asymmetric. Control accounts may cover topics not present among IO accounts since we do not exclude control content that is not matched to IO hashtag use.

All of these discrepancies might impact detection algorithms or descriptive studies aiming to differentiate IO and control accounts.

Finally, while IOs can spread across multiple social media platforms \citep{wilson2021cross}, our datasets are focused on a single platform. Considering how campaigns unfold on multiple online social networks is a valuable direction for future studies.

\section{Ethical Statement}
\label{sec:ethical}

The collection and release of the dataset was deemed exempt from review by the Indiana University IRB (protocol 1102004860) as it contains de-identified public data with minimal or no risks to the subjects. 
Although PII has been anonymized to protect privacy, sensitive information (such as political opinions, racial or ethnic origins, or religious beliefs) may still be inferred from the content of posts. If combined with other data sources, this could lead to unintended re-identification. Therefore, researchers should handle the data responsibly, ensure compliance with ethical guidelines, and refrain from attempting to de-anonymize or link the data with other datasets in a manner that could compromise individual privacy. 
Finally, we allow researchers to download the datasets in small segments to comply with platform terms. 

\section{Acknowledgments}
\label{sec:acknowledgements}

We are grateful to David Axelrod and Filipi Nascimento Silva for helpful discussions on the limitations of the datasets.
This work was supported in part by the Knight Foundation, Craig Newmark Philanthropies, and Volkswagen Foundation. The funders had no role in study design, data collection, and analysis, the decision to publish, or the preparation of the manuscript.

\section{Contributions}

OCS, MP, and ACN contributed to the conceptualization of the project, investigation, data curation, validation and analysis, and writing of the manuscript.
OCS prepared the figures. 
LY contributed to data anonymization. 
LL conceived a consistent anonymization strategy to link IO and control posts and edited the manuscript.
FM and AF contributed to the conceptualization of the project, acquired funding, supervised the project, and reviewed and edited the manuscript.

\bibliography{io_control_dataset}

\section{Ethical Checklist}

\begin{enumerate}

\item For most authors...
\begin{enumerate}
    \item  Would answering this research question advance science without violating social contracts, such as violating privacy norms, perpetuating unfair profiling, exacerbating the socio-economic divide, or implying disrespect to societies or cultures?
    \answerYes{Yes}
  \item Do your main claims in the abstract and introduction accurately reflect the paper's contributions and scope?
    \answerYes{Yes}
   \item Do you clarify how the proposed methodological approach is appropriate for the claims made? 
    \answerYes{Yes}
   \item Do you clarify what are possible artifacts in the data used, given population-specific distributions?
    \answerYes{Yes}
  \item Did you describe the limitations of your work?
    \answerYes{Yes}
  \item Did you discuss any potential negative societal impacts of your work?
    \answerYes{Yes}
      \item Did you discuss any potential misuse of your work?
    \answerYes{Yes}
    \item Did you describe steps taken to prevent or mitigate potential negative outcomes of the research, such as data and model documentation, data anonymization, responsible release, access control, and the reproducibility of findings?
    \answerYes{Yes}
  \item Have you read the ethics review guidelines and ensured that your paper conforms to them?
    \answerYes{Yes}
\end{enumerate}

\item Additionally, if you are using existing assets (e.g., code, data, models) or curating/releasing new assets...
\begin{enumerate}
  \item If your work uses existing assets, did you cite the creators?
    \answerNo{The IO data is no longer available to the public, therefore we do not name the source platform.}
  \item Did you mention the license of the assets?
    \answerYes{The license of the new datasets is indicated in the data repository.}
  \item Did you include any new assets in the supplemental material or as a URL?
    \answerYes{Yes, we provide the Zenodo URL and DOI for the datasets.}
  \item Did you discuss whether and how consent was obtained from people whose data you're using/curating?
    \answerYes{Yes, in the Ethical Statement section.}
  \item Did you discuss whether the data you are using/curating contains personally identifiable information or offensive content?
    \answerYes{Yes}
    \item If you are curating or releasing new datasets, did you discuss how you intend to make your datasets FAIR?
    \answerYes{Yes}
    \item If you are curating or releasing new datasets, did you create a Datasheet for the Dataset? 
    \answerYes{Yes}
\end{enumerate}

\end{enumerate}


\end{document}

%% file: tables/descriptive_statistics.tex
\begin{table*}
\resizebox{\textwidth}{!}{%
\begin{tabular}{r|rr|rr|rr|rr|rr|rr|rr|rr}
\multicolumn{1}{c|}{} &
  \multicolumn{2}{c|}{\textbf{Years}} &
  \multicolumn{2}{c|}{\textbf{Accounts}} &
  \multicolumn{2}{c|}{\textbf{Posts}} &
  \multicolumn{2}{c|}{\textbf{Reposts}} &
  \multicolumn{2}{c|}{\textbf{Replies}} &
  \multicolumn{2}{c|}{\textbf{Unique Hashtags}} &
  \multicolumn{2}{c|}{\textbf{Unique URLs}} &
  \multicolumn{2}{c}{\textbf{Unique Mentioned Accounts}} \\ 
\multicolumn{1}{c|}{\multirow{-2}{*}{\textbf{Campaign}}} &
  \multicolumn{1}{c}{{\textbf{IO}}} &
  \multicolumn{1}{c|}{{\textbf{Control}}} &
  \multicolumn{1}{c}{{\textbf{IO}}} &
  \multicolumn{1}{c|}{{\textbf{Control}}} &
  \multicolumn{1}{c}{{\textbf{IO}}} &
  \multicolumn{1}{c|}{{\textbf{Control}}} &
  \multicolumn{1}{c}{{\textbf{IO}}} &
  \multicolumn{1}{c|}{{\textbf{Control}}} &
  \multicolumn{1}{c}{{\textbf{IO}}} &
  \multicolumn{1}{c|}{{\textbf{Control}}} &
  \multicolumn{1}{c}{{\textbf{IO}}} &
  \multicolumn{1}{c|}{{\textbf{Control}}} &
  \multicolumn{1}{c}{{\textbf{IO}}} &
  \multicolumn{1}{c|}{{\textbf{Control}}} &
  \multicolumn{1}{c}{{\textbf{IO}}} &
  \multicolumn{1}{c}{{\textbf{Control}}} \\ \hline
  
{Armenia} & {2014-2020} & {2014-2020} & {31} & {1,767} & {72,960} & {69,118} & {522} & {35,167} & {263} & {6,253} & {866} & {26,859} & {89,989} & {58,022} & {607} & {31,127} \\

{Bangladesh} & {2009-2018} & {2010-2018} & {11} & {929} & {26,212} & {35,974} & {1,020} & {16,640} & {310} & {2,504} & {950} & {19,113} & {23,532} & {28,831} & {618} & {16,541} \\

{Catalonia} & {2011-2019} & {2011-2019} & {76} & {2,607} & {9,489} & {115,520} & {6,096} & {95,436} & {2,201} & {9,540} & {699} & {25,049} & {2,226} & {75,634} & {2,260} & {39,974} \\

{China\_1} & {2008-2019} & {2008-2019} & {699} & {42,120} & {1,898,108} & {1,772,646} & {356,054} & {1,026,404} & {178,496} & {205,897} & {25,024} & {226,561} & {553,156} & {915,030} & {139,206} & {524,836} \\

{China\_2} & {2007-2019} & {2008-2019} & {191} & {35,806} & {1,708,078} & {1,705,976} & {671,070} & {338,756} & {223,878} & {293,122} & {74,879} & {176,105} & {629,535} & {526,831} & {170,215} & {588,095} \\

{Cuba} & {2010-2020} & {2010-2020} & {503} & {30,099} & {4,802,243} & {1,353,088} & {3,341,163} & {626,977} & {151,121} & {128,785} & {88,022} & {178,692} & {1,586,187} & {777,179} & {141,107} & {391,511} \\

{Ecuador} & {2010-2019} & {2010-2019} & {787} & {21,138} & {700,240} & {770,289} & {577,749} & {456,703} & {46,159} & {67,271} & {32,283} & {145,667} & {135,890} & {483,906} & {49,739} & {237,738} \\

{Egypt\_UAE} & {2012-2019} & {2012-2019} & {240} & {370} & {214,898} & {14,712} & {71,883} & {8,164} & {11,584} & {1,211} & {33,864} & {5,174} & {155,877} & {7,690} & {10,851} & {4,972} \\

{Ghana\_Nigeria} & {2014-2020} & {2014-2020} & {60} & {1,166} & {39,964} & {43,445} & {16,310} & {30,304} & {8,697} & {4,995} & {8,340} & {15,247} & {15,130} & {29,345} & {11,168} & {28,054} \\

{Iran\_1} & {2010-2018} & {2012-2018} & {660} & {5,015} & {1,122,936} & {199,853} & {232,337} & {84,206} & {339,350} & {17,306} & {105,598} & {55,882} & {612,899} & {137,711} & {296,150} & {72,244} \\

{Iran\_2} & {2009-2020} & {2011-2020} & {389} & {6,615} & {1,302,012} & {329,781} & {502,373} & {148,585} & {30,405} & {30,822} & {90,857} & {50,394} & {928,318} & {140,738} & {44,628} & {72,115} \\

{Iran\_3} & {2011-2019} & {2014-2016} & {210} & {2,240} & {1,963,141} & {91,297} & {1,609,566} & {48,025} & {222,737} & {7,596} & {156,904} & {18,144} & {388,313} & {41,499} & {102,584} & {22,787} \\

{Iran\_4} & {2008-2019} & {2013-2019} & {2,519} & {7,612} & {254,781} & {306,804} & {90,324} & {181,129} & {62,800} & {27,974} & {33,806} & {76,313} & {76,287} & {201,689} & {37,031} & {105,735} \\

{Iran\_5} & {2020-2020} & {2020-2020} & {104} & {1,247} & {2,450} & {42,955} & {551} & {30,996} & {663} & {5,706} & {148} & {14,280} & {1,166} & {27,725} & {1,024} & {28,314} \\

{Iran\_6} & {2009-2020} & {2010-2020} & {209} & {16,885} & {560,571} & {729,241} & {100,446} & {467,772} & {89,139} & {61,339} & {37,964} & {147,664} & {462,034} & {554,717} & {56,718} & {240,109} \\

{Qatar} & {2010-2020} & {2013-2020} & {29} & {19,481} & {220,254} & {805,934} & {163,470} & {548,761} & {5,647} & {70,332} & {16,347} & {125,488} & {27,757} & {385,300} & {23,036} & {162,002} \\

{Russia\_1} & {2009-2018} & {2009-2018} & {3,293} & {31,303} & {1,826,344} & {1,823,859} & {697,697} & {406,516} & {49,039} & {138,448} & {60,876} & {176,935} & {1,115,538} & {910,924} & {188,637} & {342,223} \\

{Russia\_2} & {2010-2018} & {2011-2018} & {361} & {2,175} & {920,761} & {90,291} & {712,800} & {36,534} & {72,380} & {9,968} & {82,946} & {25,553} & {460,929} & {51,324} & {137,222} & {36,283} \\

{Russia\_3} & {2019-2020} & {2020-2020} & {5} & {780} & {1,368} & {31,741} & {494} & {23,135} & {72} & {4,572} & {205} & {9,238} & {803} & {19,129} & {242} & {17,770} \\

{Russia\_4} & {2009-2020} & {2010-2020} & {24} & {36,533} & {68,914} & {1,776,226} & {21,775} & {976,712} & {5,587} & {168,996} & {9,439} & {322,534} & {42,268} & {1,251,577} & {8,118} & {485,809} \\

{Russia\_5} & {2013-2020} & {2014-2020} & {51} & {13,077} & {26,684} & {683,219} & {3,746} & {443,987} & {2,870} & {67,512} & {6,640} & {163,071} & {21,308} & {546,679} & {3,534} & {225,216} \\

{Spain} & {2019-2019} & {2019-2019} & {216} & {1,681} & {56,712} & {62,778} & {27,042} & {50,549} & {8,985} & {5,589} & {1,413} & {15,792} & {10,670} & {41,796} & {3,558} & {33,268} \\

{Thailand} & {2015-2020} & {2018-2020} & {455} & {2,549} & {21,385} & {121,738} & {11,199} & {109,226} & {6,844} & {5,157} & {1,615} & {17,797} & {2,871} & {66,251} & {1,432} & {36,824} \\

{UAE} & {2011-2019} & {2011-2019} & {3,658} & {10,545} & {1,325,530} & {382,262} & {644,486} & {233,446} & {99,152} & {38,579} & {92,920} & {94,113} & {402,945} & {229,095} & {105,744} & {125,456} \\

{Venezuela\_1} & {2015-2018} & {2016-2018} & {578} & {6,327} & {984,980} & {261,310} & {1,416} & {164,960} & {5} & {31,784} & {12,472} & {41,680} & {1,645,110} & {182,546} & {305} & {87,065} \\

{Venezuela\_2} & {2012-2019} & {2012-2019} & {33} & {3,865} & {569,453} & {156,989} & {28,959} & {99,952} & {125} & {18,190} & {930} & {33,510} & {892,852} & {113,081} & {563} & {62,706} \\

\hline

\end{tabular}%
}
\caption{Descriptive Statistics. For IO and control data in each dataset, we report the time frame, numbers of accounts, posts, replies, reshares, unique hashtags, URLs, and mentioned accounts.}
\label{tab:descriptive-stats}
\end{table*}

%% file: io_control_dataset.bbl
\begin{thebibliography}{44}
\providecommand{\natexlab}[1]{#1}

\bibitem[{Addawood et~al.(2019)Addawood, Badawy, Lerman, and Ferrara}]{addawood2019linguistic}
Addawood, A.; Badawy, A.; Lerman, K.; and Ferrara, E. 2019.
\newblock {Linguistic Cues to Deception: Identifying Political Trolls on Social Media}.
\newblock In \emph{Proc. Intl. AAAI Conf. on Web and Social Media}, volume~13, 15--25.

\bibitem[{Alizadeh et~al.(2020)Alizadeh, Shapiro, Buntain, and Tucker}]{alizadeh2020content-based}
Alizadeh, M.; Shapiro, J.~N.; Buntain, C.; and Tucker, J.~A. 2020.
\newblock {Content-based Features Predict Social Media Influence Operations}.
\newblock \emph{Science Advances}, 6(30).

\bibitem[{Badawy et~al.(2019)Badawy, Addawood, Lerman, and Ferrara}]{badawy2019characterizing}
Badawy, A.; Addawood, A.; Lerman, K.; and Ferrara, E. 2019.
\newblock {Characterizing the 2016 Russian IRA Influence Campaign}.
\newblock \emph{Social Network Analysis and Mining}, 9: 1--11.

\bibitem[{Bradshaw and Howard(2017)}]{bradshaw2017troops}
Bradshaw, S.; and Howard, P. 2017.
\newblock {Troops, Trolls and Troublemakers: A Global Inventory of Organized Social Media Manipulation}.
\newblock Technical report, University of Oxford.

\bibitem[{Cima et~al.(2024)Cima, Mannocci, Avvenuti, Tesconi, and Cresci}]{cima2024coordinated}
Cima, L.; Mannocci, L.; Avvenuti, M.; Tesconi, M.; and Cresci, S. 2024.
\newblock {Coordinated behavior in information operations on Twitter}.
\newblock \emph{IEEE Access}.

\bibitem[{Elmas(2023)}]{elmas2023analyzing}
Elmas, T. 2023.
\newblock {Analyzing Activity and Suspension Patterns of Twitter Bots Attacking Turkish Twitter Trends by a Longitudinal Dataset}.
\newblock In \emph{Companion Proc. ACM Web Conf.}, 1404--1412.

\bibitem[{Ezzeddine et~al.(2023)Ezzeddine, Ayoub, Giordano, Nogara, Sbeity, Ferrara, and Luceri}]{ezzeddine2023exposing}
Ezzeddine, F.; Ayoub, O.; Giordano, S.; Nogara, G.; Sbeity, I.; Ferrara, E.; and Luceri, L. 2023.
\newblock {Exposing Influence Campaigns in the Age of LLMs: A Behavioral-based AI Approach to Detecting State-Sponsored Trolls}.
\newblock \emph{EPJ Data Science}, 12(1): 46.

\bibitem[{Facebook(2021)}]{facebook2021threatreport}
Facebook. 2021.
\newblock {Threat Report: The State of Influence Operations 2017-2020}.
\newblock Technical report, Facebook.

\bibitem[{Fecher et~al.(2022)Fecher, Reich, Taylor, and Warren}]{fecher2022oh}
Fecher, L.; Reich, T.; Taylor, J.; and Warren, P. 2022.
\newblock {Oh, The Places You'll Guo! The Tactics and Impact of a Chinese Multilingual Narrative Flooding Campaign through Political Cartoons}.
\newblock \url{https://open.clemson.edu/cgi/viewcontent.cgi?article=1000&context=mfh_ci_reports}.

\bibitem[{Guo and Vosoughi(2022)}]{guo2022large}
Guo, X.; and Vosoughi, S. 2022.
\newblock {A large-scale longitudinal multimodal dataset of state-backed information operations on Twitter}.
\newblock In \emph{Proc. of the Intl. AAAI Conf. on Web and Social Media}, volume~16, 1245--1250.

\bibitem[{Im et~al.(2020)Im, Chandrasekharan, Sargent, Lighthammer, Denby, Bhargava, Hemphill, Jurgens, and Gilbert}]{im2020still}
Im, J.; Chandrasekharan, E.; Sargent, J.; Lighthammer, P.; Denby, T.; Bhargava, A.; Hemphill, L.; Jurgens, D.; and Gilbert, E. 2020.
\newblock {Still Out There: Modeling and Identifying Russian Troll Accounts on Twitter}.
\newblock In \emph{12th ACM Conf. on Web Science}.

\bibitem[{Keller et~al.(2020)Keller, Schoch, Stier, and Yang}]{keller2020political}
Keller, F.~B.; Schoch, D.; Stier, S.; and Yang, J. 2020.
\newblock {Political Astroturfing on Twitter: How to Coordinate a Disinformation Campaign}.
\newblock \emph{Political communication}.

\bibitem[{Kong et~al.(2023)Kong, Calderon, Ram, Boichak, and Rizoiu}]{kong23Interval-censored}
Kong, Q.; Calderon, P.; Ram, R.; Boichak, O.; and Rizoiu, M.-A. 2023.
\newblock {Interval-censored Transformer Hawkes: Detecting Information Operations using the Reaction of Social Systems}.
\newblock In \emph{Proc. ACM Web Conf.}, 1813–--1821.

\bibitem[{Lazer et~al.(2018)Lazer, Baum, Benkler, Berinsky, Greenhill, Menczer, Metzger, Nyhan, Pennycook, Rothschild, Schudson, Sloman, Sunstein, Thorson, Watts, and Zittrain}]{Lazer-fake-news-2018}
Lazer, D.; Baum, M.; Benkler, Y.; Berinsky, A.; Greenhill, K.; Menczer, F.; Metzger, M.; Nyhan, B.; Pennycook, G.; Rothschild, D.; Schudson, M.; Sloman, S.; Sunstein, C.; Thorson, E.; Watts, D.; and Zittrain, J. 2018.
\newblock The science of fake news.
\newblock \emph{Science}, 359(6380): 1094--1096.

\bibitem[{Linvill and Warren(2020)}]{linvill_warren_2020}
Linvill, D.~L.; and Warren, P.~L. 2020.
\newblock {Engaging with Others: How the IRA Coordinated Information Operation Made Friends}.
\newblock \emph{Misinformation Review}.

\bibitem[{Luceri, Boniardi, and Ferrara(2024)}]{luceri2023leveraging}
Luceri, L.; Boniardi, E.; and Ferrara, E. 2024.
\newblock Leveraging Large Language Models to Detect Influence Campaigns in Social Media.
\newblock In \emph{Companion Proc. of the 2024 ACM Web Conf.}

\bibitem[{Luceri, Giordano, and Ferrara(2020)}]{luceri2020detecting}
Luceri, L.; Giordano, S.; and Ferrara, E. 2020.
\newblock {Detecting Troll Behavior via Inverse Reinforcement Learning: A Case Study of Russian Trolls in the 2016 US Election}.
\newblock In \emph{Intl. AAAI Conf. on Web and Social Media}, volume~14, 417--427.

\bibitem[{Luceri et~al.(2024)Luceri, Pantè, Burghardt, and Ferrara}]{luceri2024unmasking}
Luceri, L.; Pantè, V.; Burghardt, K.; and Ferrara, E. 2024.
\newblock Unmasking the web of deceit: Uncovering coordinated activity to expose information operations on Twitter.
\newblock In \emph{Proc. of the 2024 ACM Web Conf.}

\bibitem[{Merhi, Rajtmajer, and Lee(2023)}]{Merhi2021}
Merhi, M.; Rajtmajer, S.; and Lee, D. 2023.
\newblock {Information Operations in Turkey: Manufacturing Resilience with Free Twitter Accounts}.
\newblock In \emph{Proc. Intl. AAAI Conf. on Web and Social Media}.

\bibitem[{Murtfeldt et~al.(2024)Murtfeldt, Alterman, Kahveci, and West}]{murtfeldt2024rip}
Murtfeldt, R.; Alterman, N.; Kahveci, I.; and West, J.~D. 2024.
\newblock RIP Twitter API: A eulogy to its vast research contributions.
\newblock \emph{arXiv preprint arXiv:2404.07340}.

\bibitem[{Ng, Moffitt, and Carley(2022)}]{ng2022coordinated}
Ng, L. H.~X.; Moffitt, J.; and Carley, K.~M. 2022.
\newblock {Coordinated through a Web of Images: Analysis of Image-based Influence Operations from China, Iran, Russia, and Venezuela}.
\newblock \emph{Preprint arXiv:2206.03576}.

\bibitem[{Nwala, Flammini, and Menczer(2023)}]{nwala2023language}
Nwala, A.~C.; Flammini, A.; and Menczer, F. 2023.
\newblock {A Language Framework for Modeling Social Media Account Behavior}.
\newblock \emph{EPJ Data Science}, 12(1): 33.

\bibitem[{Ong and Caba{\~n}es(2018)}]{ong2018architects}
Ong, J.~C.; and Caba{\~n}es, J. V.~A. 2018.
\newblock {Architects of Networked Disinformation: Behind the Scenes of Troll Accounts and Fake News Production in the Philippines}.
\newblock Tech. rep., UMass Amherst.
\newblock \url{https://doi.org/10.7275/2cq4-5396}.

\bibitem[{Pacheco et~al.(2021)Pacheco, Hui, Torres-Lugo, Truong, Flammini, and Menczer}]{pacheco2021uncovering}
Pacheco, D.; Hui, P.-M.; Torres-Lugo, C.; Truong, B.~T.; Flammini, A.; and Menczer, F. 2021.
\newblock {Uncovering Coordinated Networks on Social Media: Methods and Case Studies.}
\newblock In \emph{Proc. Intl. AAAI Conf. on Web and Social Media}, volume~21, 455--466.

\bibitem[{Rowett(2018)}]{rowett2018strategic}
Rowett, G. 2018.
\newblock {The Strategic Need to Understand Online Memes and Modern Information Warfare Theory}.
\newblock In \emph{IEEE Big Data}.

\bibitem[{Saeed et~al.(2024)Saeed, Ali, Paudel, Blackburn, and Stringhini}]{saeed2024unraveling}
Saeed, M.~H.; Ali, S.; Paudel, P.; Blackburn, J.; and Stringhini, G. 2024.
\newblock {Unraveling the Web of Disinformation: Exploring the Larger Context of State-Sponsored Influence Campaigns on Twitter}.
\newblock \emph{arXiv preprint arXiv:2407.18098}.

\bibitem[{{Senate Select Committee on Intelligence}(2019)}]{senate_intelligence_committee_2019}
{Senate Select Committee on Intelligence}. 2019.
\newblock {Report of the Select Committee on Intelligence United States Senate on Russian Active Measures Campaigns and Interference in the 2016 U.S. Election Volume 2: Russia's Use of Social Media with Additional Views}.
\newblock \url{https://www.intelligence.senate.gov/sites/default/files/documents/Report_Volume2.pdf}.
\newblock Accessed: 2024-01-16.

\bibitem[{Shao et~al.(2018)Shao, Ciampaglia, Varol, Yang, Flammini, and Menczer}]{Shao18hoaxybots}
Shao, C.; Ciampaglia, G.~L.; Varol, O.; Yang, K.; Flammini, A.; and Menczer, F. 2018.
\newblock {The Spread of Low-credibility Content by Social bots}.
\newblock \emph{Nature Communications}, 9: 4787.

\bibitem[{Sharma et~al.(2021)Sharma, Zhang, Ferrara, and Liu}]{sharm2021hidden}
Sharma, K.; Zhang, Y.; Ferrara, E.; and Liu, Y. 2021.
\newblock {Identifying Coordinated Accounts on Social Media through Hidden Influence and Group Behaviours}.
\newblock In \emph{27th ACM SIGKDD Conf. on Knowledge Discovery \& Data Mining}, 1441--1451.

\bibitem[{Smith, Ehrett, and Warren(2024)}]{smith2024unsupervised}
Smith, D.~H.; Ehrett, C.; and Warren, P.~L. 2024.
\newblock {Unsupervised Detection of Coordinated Information Operations in the Wild}.
\newblock \emph{arXiv preprint arXiv:2401.06205}.

\bibitem[{{Stanford Internet Observatory}(2021)}]{sio}
{Stanford Internet Observatory}. 2021.
\newblock Published reports of the Stanford Internet Observatory.
\newblock \url{https://github.com/stanfordio/publications}.

\bibitem[{Starbird(2019)}]{starbird2019disinformation}
Starbird, K. 2019.
\newblock Disinformation's spread: bots, trolls and all of us.
\newblock \emph{Nature}, 571(7766): 449--450.

\bibitem[{Stewart, Arif, and Starbird(2018)}]{stewart2018examining}
Stewart, L.~G.; Arif, A.; and Starbird, K. 2018.
\newblock {Examining Trolls and Polarization with a Retweet Network}.
\newblock In \emph{Proc. ACM WSDM Workshop on Misinformation and Misbehavior Mining on the Web (MIS2)}.
\newblock \url{https://api.semanticscholar.org/CorpusID:44033303}.

\bibitem[{Torres-Lugo et~al.(2022)Torres-Lugo, Pote, Nwala, and Menczer}]{Torres2022deletions}
Torres-Lugo, C.; Pote, M.; Nwala, A.; and Menczer, F. 2022.
\newblock {Manipulating Twitter through Deletions}.
\newblock In \emph{Proc. Intl. AAAI Conf. on Web and Social Media}, 1029--1039.

\bibitem[{Uyheng, Cruickshank, and Carley(2022)}]{uyheng2022mapping}
Uyheng, J.; Cruickshank, I.~J.; and Carley, K.~M. 2022.
\newblock {Mapping State-sponsored Information Operations with Multi-view Modularity Clustering}.
\newblock \emph{EPJ Data Science}, 11(1): 25.

\bibitem[{Vargas, Emami, and Traynor(2020)}]{vargas2020detection}
Vargas, L.; Emami, P.; and Traynor, P. 2020.
\newblock {On the Detection of Disinformation Campaign Activity with Network Analysis}.
\newblock In \emph{Proc. of the 2020 ACM SIGSAC Conf. on Cloud Computing Security Workshop}, 133--146.

\bibitem[{Vosoughi, Roy, and Aral(2018)}]{vosoughi2018spread}
Vosoughi, S.; Roy, D.; and Aral, S. 2018.
\newblock {The Spread of True and False News Online}.
\newblock \emph{Science}.

\bibitem[{Wilson and Starbird(2021)}]{wilson2021cross}
Wilson, T.; and Starbird, K. 2021.
\newblock {Cross-platform Information Operations: Mobilizing Narratives \& Building Resilience through both'Big'\&'Alt'Tech}.
\newblock \emph{Proc. of the ACM on Human-Computer Interaction}.

\bibitem[{Woolley and Howard(2018{\natexlab{a}})}]{woolley2018computational}
Woolley, S.~C.; and Howard, P.~N. 2018{\natexlab{a}}.
\newblock \emph{{Computational Propaganda: Political Parties, Politicians, and Political Manipulation on Social Media}}.
\newblock Oxford University Press.

\bibitem[{Woolley and Howard(2018{\natexlab{b}})}]{woolley2019conclusion}
Woolley, S.~C.; and Howard, P.~N. 2018{\natexlab{b}}.
\newblock {Conclusion: Political Parties, Politicians, and Computational Propaganda}.
\newblock In \emph{Computational propaganda: Political parties, politicians, and political manipulation on social media}, 241--248. Oxford University Press.

\bibitem[{Yang and Menczer(2024)}]{yang2023anatomy}
Yang, K.-C.; and Menczer, F. 2024.
\newblock {Anatomy of an AI-powered Malicious Social Botnet}.
\newblock \emph{Journal of Quantitative Description: Digital Media}, 4.

\bibitem[{Zannettou et~al.(2020)Zannettou, Caulfield, Bradlyn, De~Cristofaro, Stringhini, and Blackburn}]{zannettou2020characterizing}
Zannettou, S.; Caulfield, T.; Bradlyn, B.; De~Cristofaro, E.; Stringhini, G.; and Blackburn, J. 2020.
\newblock Characterizing the Use of Images in State-Sponsored Information Warfare Operations by Russian Trolls on Twitter.
\newblock \emph{Proc. of the Intl. AAAI Conf. on Web and Social Media}.

\bibitem[{Zannettou et~al.(2019{\natexlab{a}})Zannettou, Caulfield, De~Cristofaro, Sirivianos, Stringhini, and Blackburn}]{zannettou2019disinformation}
Zannettou, S.; Caulfield, T.; De~Cristofaro, E.; Sirivianos, M.; Stringhini, G.; and Blackburn, J. 2019{\natexlab{a}}.
\newblock {Disinformation Warfare: Understanding State-sponsored Trolls on Twitter and their Influence on the Web}.
\newblock In \emph{Companion Proc. WWW Conf.}, 218--226.

\bibitem[{Zannettou et~al.(2019{\natexlab{b}})Zannettou, Caulfield, Setzer, Sirivianos, Stringhini, and Blackburn}]{zannettou2019let}
Zannettou, S.; Caulfield, T.; Setzer, W.; Sirivianos, M.; Stringhini, G.; and Blackburn, J. 2019{\natexlab{b}}.
\newblock {Who Let the Trolls Out? Towards Understanding State-sponsored Trolls}.
\newblock In \emph{Proc. 10th ACM Conf. on Web Science}, 353--362.

\end{thebibliography}
